\documentclass[%
 reprint,
superscriptaddress,
 amsmath,amssymb,
 aps,
 showkeys,
 nolongbibliography,
prb,
]{revtex4-2}

\usepackage{graphicx}
\usepackage{dcolumn}
\usepackage{bm}
\usepackage{float}
\usepackage{lipsum, babel}

\usepackage[table]{xcolor}
\usepackage{tikz}

\usepackage{filecontents}

\begin{document}

\preprint{APS/123-QED}

\title{Superconductivity in the doped polymerized fullerite clathrate from first principles}

\author{Jorge Laranjeira}
 \email{jorgelaranjeira@ua.pt}
 \affiliation{Departamento de Física and CICECO, Universidade de Aveiro, 3810-193 Aveiro, Portugal}
 
\author{Ion Errea}%
\affiliation{Fisika Aplikatua Saila, Gipuzkoako Ingeniaritza Eskola, University of the Basque Country (UPV/EHU), Europa Plaza 1, 20018 Donostia/San Sebastián, Spain}
\affiliation{Centro de Física de Materiales (CSIC-UPV/EHU), Manuel de Lardizabal Pasealekua 5, 20018 Donostia/San Sebastián, Spain}
\affiliation{Donostia International Physics Center (DIPC), Manuel de Lardizabal Pasealekua 4, 20018 Donostia/San Sebastián, Spain}

\author{\DJ or\dj e    Dangi\'{c}}%
\affiliation{Fisika Aplikatua Saila, Gipuzkoako Ingeniaritza Eskola, University of the Basque Country (UPV/EHU), Europa Plaza 1, 20018 Donostia/San Sebastián, Spain}
\affiliation{Centro de Física de Materiales (CSIC-UPV/EHU), Manuel de Lardizabal Pasealekua 5, 20018 Donostia/San Sebastián, Spain}

\author{Leonel Marques}%
\affiliation{Departamento de Física and CICECO, Universidade de Aveiro, 3810-193 Aveiro, Portugal}%

\author{Manuel Melle-Franco}
\affiliation{Departamento de Química and CICECO, Universidade de Aveiro, 3810-193 Aveiro, Portugal}%

\author{Karol Struty\'nski}
\affiliation{Departamento de Química and CICECO, Universidade de Aveiro, 3810-193 Aveiro, Portugal}%

\date{\today}

\begin{abstract}
Superconductivity in the new polymerized $\mathrm{C}_{60}$ fullerite clathrate doped with simple metals was investigated through density functional theory methods. The phonon dispersion curves were systematically computed for the clathrate structures containing the guest dopants (Li, Na, K, Be, Mg, Ca, Al, Ga, Ge) in one of the two distinct cages, fullerene-like $\mathrm{C}_{60}$ and sodalite-like $\mathrm{C}_{24}$, exhibited by this structure. Only four of these structures, Li@$\mathrm{C}_{24}$, Na@$\mathrm{C}_{24}$, Ga@$\mathrm{C}_{24}$ and Be@$\mathrm{C}_{60}$, are dynamically stable in the harmonic approximation. They all show superconducting behavior, although their critical temperatures are predicted to be below 2 K. 		
\end{abstract}

\maketitle

\section*{Introduction}

The achievement of room temperature superconductivity is extremely desirable for technological applications \cite{revIon}. Very high superconducting critical temperature ($\mathrm{T}_{c}$), $\sim$200 K, was initially predicted from first principles density functional theory (DFT) for the $\mathrm{H}_{3}\mathrm{S}$ compound \cite{hs3_theoric2014}, soon followed by its experimental observation \cite{hs3_experimental2015} under a pressure of 150 GPa. These exciting results fostered huge efforts, led again by computational methods, to find new materials with even higher $\mathrm{T}_{c}$ by exploring hydrogen bearing compounds. A new lanthanum hydride, LaH$_{10}$, in which the hydrogen atoms form sodalite-like cages, was found at 140-220 GPa and displaying T$_c$ of approximately 250 K \cite{lah10_exp,drozdov2019}. This near-room temperature superconducting phase is only stable under megabar pressures and cannot be quenched to room conditions, thus precluding practical applications. The focus of investigations is now directed to the search for superconducting materials that can operate at room pressure. One strategy to attain this goal is to consider other types of clathrate structures that still use a light element in its building blocks, a known key ingredient for high $\mathrm{T}_{c}$ conventional superconductivity \cite{revIon}, while being able to retain the clathrate structure at ambient pressure. Carbon clathrates are notable candidates since, besides containing a light element such as carbon, they exhibit strong covalent bonds that lead to large phonon frequencies, which is another key ingredient for high $\mathrm{T}_{c}$ superconductivity \cite{dicataldo,sodalite}. 

Unfortunately, experimental carbon clathrates are yet to be synthesized, although an important step towards its realization was done recently with the high-pressure synthesis of mixed carbon-boron clathrates, containing guest metallic atoms \cite{strobel2020,Zhu2020}. Initial predictions based on DFT methods were again crucial to guide the investigations towards their experimental realization \cite{tao2015}. The mixed carbon-boron clathrate doped with strontium guest-atoms was hypothesized as a potential candidate for phonon-mediated superconductivity at ambient conditions, since its electronic structure showed the coexistence of steep and flat bands close to the Fermi level \cite{Zhu2020}. Indeed ab initio DFT simulations performed on the strontium and barium doped carbon-boron clathrates yield T$_c$'s of 40 K and 43 K, respectively, at room pressure \cite{srbac3b3}. Furthermore, an increase in T$_c$ up to 88 K was predicted in this class of compounds for the potassium-lead binary-guest clathrate \cite{dicataldo,nisha_zurek}. 

A significant number of superconducting doped carbon allotropes have already been synthesized. Boron doped diamond yields a $\mathrm{T}_{c}$ of 4 K for a doping level of 2.5\% \cite{ekimov_bdiamond} and 11.4 K for a doping level of 5\% \cite{Bdopeddiamond}. Graphite intercalation compounds such as Yb$\mathrm{C}_{6}$ and Ca$\mathrm{C}_{6}$, display $\mathrm{T}_{c}$'s of 6.5 K and 11.5 K, respectively \cite{Yc6}. Also, the alkali-metal fulleride RbCs$_2\mathrm{C}_{60}$ displays a $\mathrm{T}_{c}$ of 33 K at room pressure. The compound Cs$_3\mathrm{C}_{60}$ presents a higher $\mathrm{T}_{c}$ 40 K, nevertheless, it requires a pressure of 1.5 GPa to stabilize the cubic superconducting phase \cite{gunnarsson,PALSTRA}. A theoretical study of a sodalite-like carbon clathrate, with a structure identical to that of the mixed carbon-boron clathrate, revealed that it could achieve high $\mathrm{T}_{c}$, 116 K, at room pressure when doped with sodium \cite{sodalite}.

Very recently, we have proposed a new carbon clathrate based on the polymerized fullerite, which is expected to be synthesized by subjecting fullerite $\mathrm{C}_{60}$ to high-pressure and high-temperature (HPHT) treatment. Some experimental high-pressure $\mathrm{C}_{60}$ phases found in the literature have interfullerene distances identical to that of the polymerized fullerite clathrate structure, showing evidence that it could be synthesized once the proper thermodynamic path is found \cite{LARANJEIRA2022}. Interestingly, its electronic structure has a narrow bandgap of 0.68 eV and its density of states (DOS), shown in Figure S1 of the Supporting  Information  (SI), displays a strong peak above ($\sim$0.7 eV) the Fermi level. We found that this peak is due to the three nearly-flat $\pi ^*$ bands from the remaining sp$^2$ carbons. The new clathrate also presents large phonon frequencies with the highest mode lying at $\sim$1350 cm$^{-1}$, which suggests that it could be a good candidate for displaying superconductivity once properly doped. Here we explore this possibility by performing a systematic DFT study of the doping effect on the fullerite $\mathrm{C}_{60}$ clathrate structure. Nine simple metals were tested as guest-atoms being inserted in one of the two cages, fullerene-like $\mathrm{C}_{60}$ and sodalite-like $\mathrm{C}_{24}$, exhibited by this clathrate structure. We compute the electronic structure and phonon dispersion curves and find that only four doped fullerite clathrates are dynamically stable, all of them showing metallic behavior. In addition, they display superconductivity but their $\mathrm{T}_{c}$'s, computed using Migdal-Eliashberg theory, are predicted to be quite low.

\section*{Methods}

For each distinct doped fullerite clathrate we optimized its structure using the Perdew-Burke-Ernzerhof (PBE) exchange correlation functional \cite{i2,i3} as implemented in the ultrasoft pseudopotentials provided with the Quantum Espresso (QE) package \cite{qe_Gionnozzi_2009,qe_Giannozzi_2017}. A k-space grid of $8\times8\times8$ was used and the self-consistent-field cycle was stopped once the energy difference between consecutive steps was smaller than $10^{-8}$ Ry. We also used a Methfessel-Paxton \cite{mp} smearing of 0.02 Ry, a kinetic energy cutoff of 60 Ry and a charge density cutoff of 600 Ry. Phonon dispersion and electron-phonon coupling calculations were performed with Density Functional Perturbation Theory (DFPT) \cite{baroni2001phonons} as implemented in QE. A q-grid of $2\times2\times2$ was used for the phonon dispersion calculation, when stability was found the q-space mesh was increased to $4\times4\times4$ prior to the electron-phonon matrix elements calculation. These parameters were converged within 1 cm$^{-1}$. To calculate the electron-phonon matrix elements we have used a smearing of 0.008 Ry for the double Dirac delta over the Fermi surface and a k-space mesh of $24\times24\times24$. To compute $\mathrm{T}_{c}$, we solved isotropic Eliashberg equations \cite{eliashberg}. We also evaluated $\mathrm{T}_{c}$ with the Allen-Dynes modified McMillan equation \cite{dynes}. 


 \section*{Results and Discussion}
 
\begin{figure}[ht]
	\centering
	\includegraphics[scale=0.1]{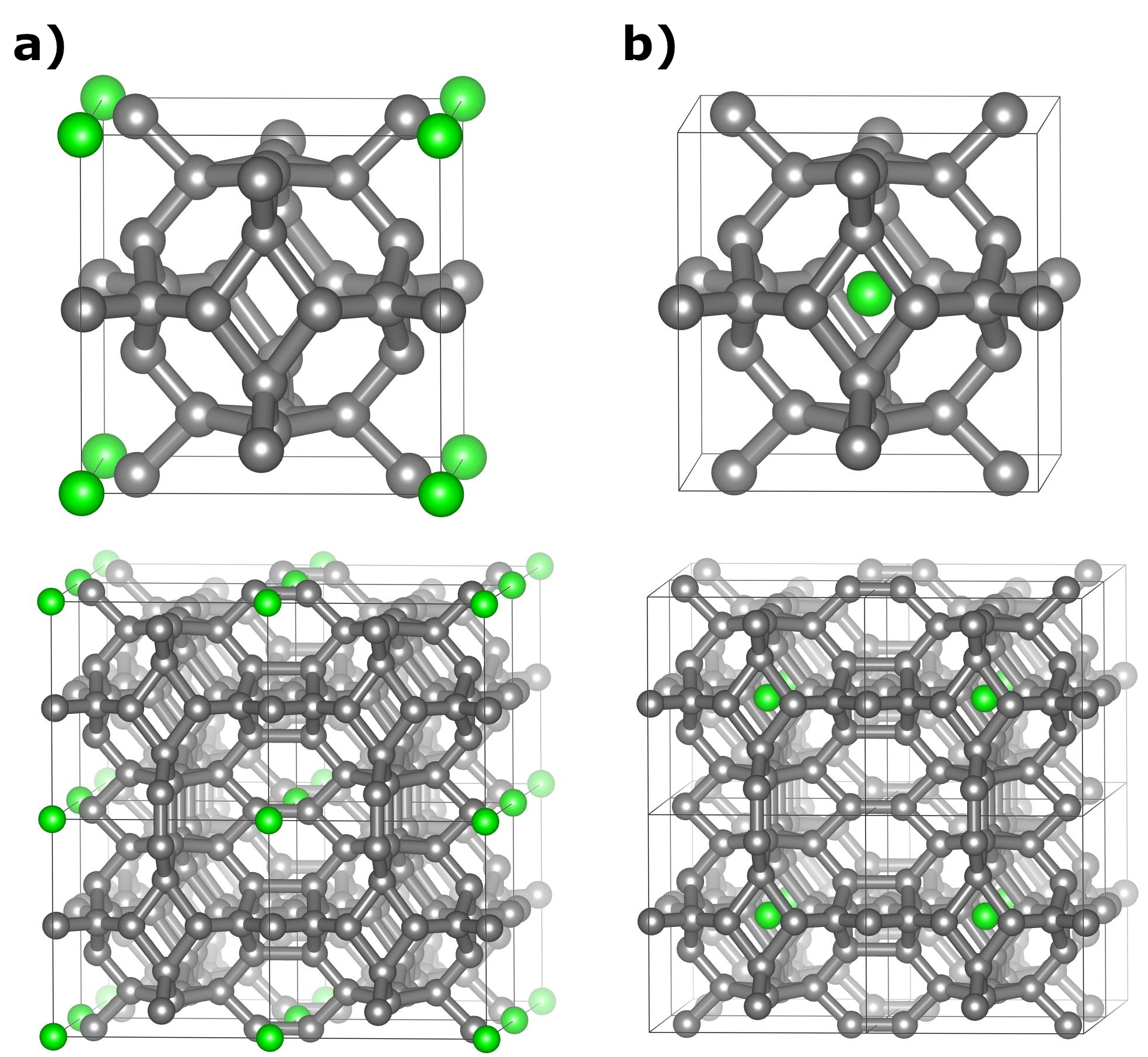}
	\caption{Doped fullerite clathrate unit cells (top panels) and $2\times2\times2$ supercell depicting the $\mathrm{C}_{60}$ cages (bottom panels) with guest atoms at: a) $\mathrm{C}_{60}$ cages, 1a Wyckoff positions; b) $\mathrm{C}_{24}$ cages, the 1b Wyckoff positions. The grey atoms are carbons and the green ones are the dopants.}
	\label{fig0}
\end{figure}

 The fullerite clathrate has a simple cubic cell, with a lattice constant of 6.21 \AA, and belongs to the Pm$\bar{3}$ space group. The structure has three independent atoms, one at 6f and two at 12k Wyckoff positions \cite{LARANJEIRA2022}. This structure exhibits two different cages, sodalite-like C$_{24}$ and fullerene-like C$_{60}$. Thus, doping may be achieved by inserting guest atoms in either one or in both cages. Here we only studied the effect of doping in only one cage. Figure \ref{fig0} a) shows the clathrate structure resulting from doping on the 1a Wyckoff position at the center of the $\mathrm{C}_{60}$ cage, hereafter denoted M@$\mathrm{C}_{60}$ (M = guest atom), while Figure \ref{fig0} b) shows the clathrate structure resulting from doping on the 1b Wyckoff position at the center of the $\mathrm{C}_{24}$ cage and hereafter denoted M@$\mathrm{C}_{24}$. Note that the resulting stoichiometry, M@$\mathrm{C}_{30}$, is the same for both insertion types. The optimized lattice constants of the differently doped structures and their relative stabilities are given in Tables S1 and S2. 

\begin{figure*}[ht!]
	\centering
	\includegraphics[scale=0.68]{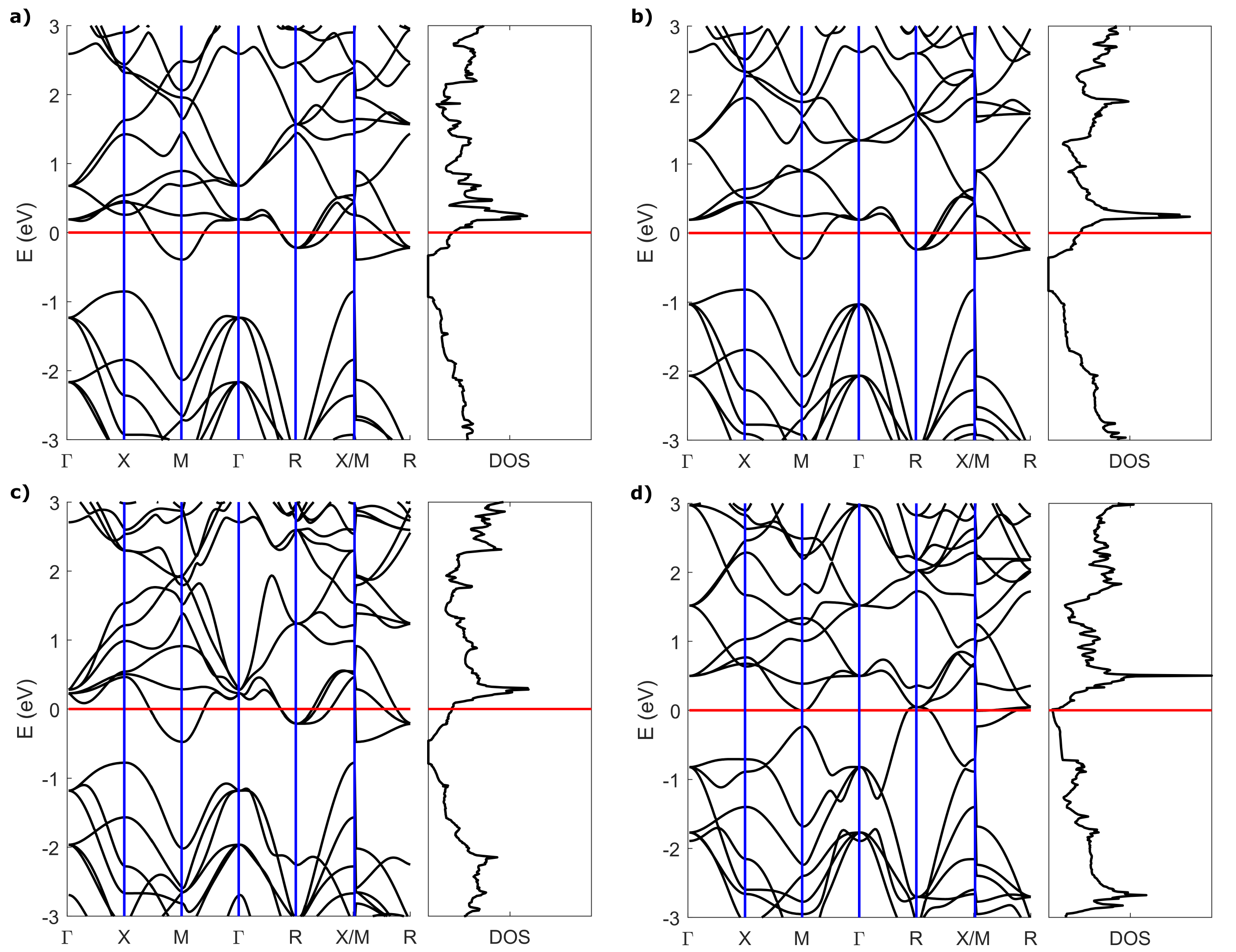}
	\caption{Electronic band structure (left panels) and density of states (right panels) for the stable compounds: a) Li@$\mathrm{C}_{24}$; b) Na@$\mathrm{C}_{24}$; c) Ga@$\mathrm{C}_{24}$; d) Be@$\mathrm{C}_{60}$. The red line indicates the Fermi level.}
	\label{fig_lamb}
\end{figure*}  
 
In an attempt to push the Fermi level to the strong DOS peak observed in the conduction band that arises from the three nearly-flat $\pi^*$ bands, we initially considered doping with trivalent ions, yttrium and scandium, at the $\mathrm{C}_{60}$ cages, thus forming the Y@$\mathrm{C}_{60}$ and Sc@$\mathrm{C}_{60}$ clathrates. Inserting these guest atoms keeps the valence and conduction bands almost unchanged, while the Fermi level moves almost to the top of the DOS peak (see the left and middle panels of Figure S2). This rigid-band picture indicates a nearly-complete charge transfer from the metal guest to the carbon framework. The number of states at the Fermi level is around 0.3 and 0.2 states/eV/spin/atom for Sc@$\mathrm{C}_{60}$ and Y@$\mathrm{C}_{60}$ respectively, a significant number that is an important prerequisite for the observation of superconductivity. Unfortunately, filling the antibonding $\pi^*$ states destabilizes the carbon framework and leads to large imaginary frequencies in the phonon dispersion curves, indicating that these doped clathrates are dynamically unstable (see the right panels of Figure S2). Moreover, we also found that they have spin-polarized ground states. The dynamical instabilities are an indication that these systems are prone to the formation of a charge-density wave that lowers the DOS at the Fermi level and prevents a magnetic ground state.

\begin{figure*}[ht!]
	\centering
	\includegraphics[scale=0.68]{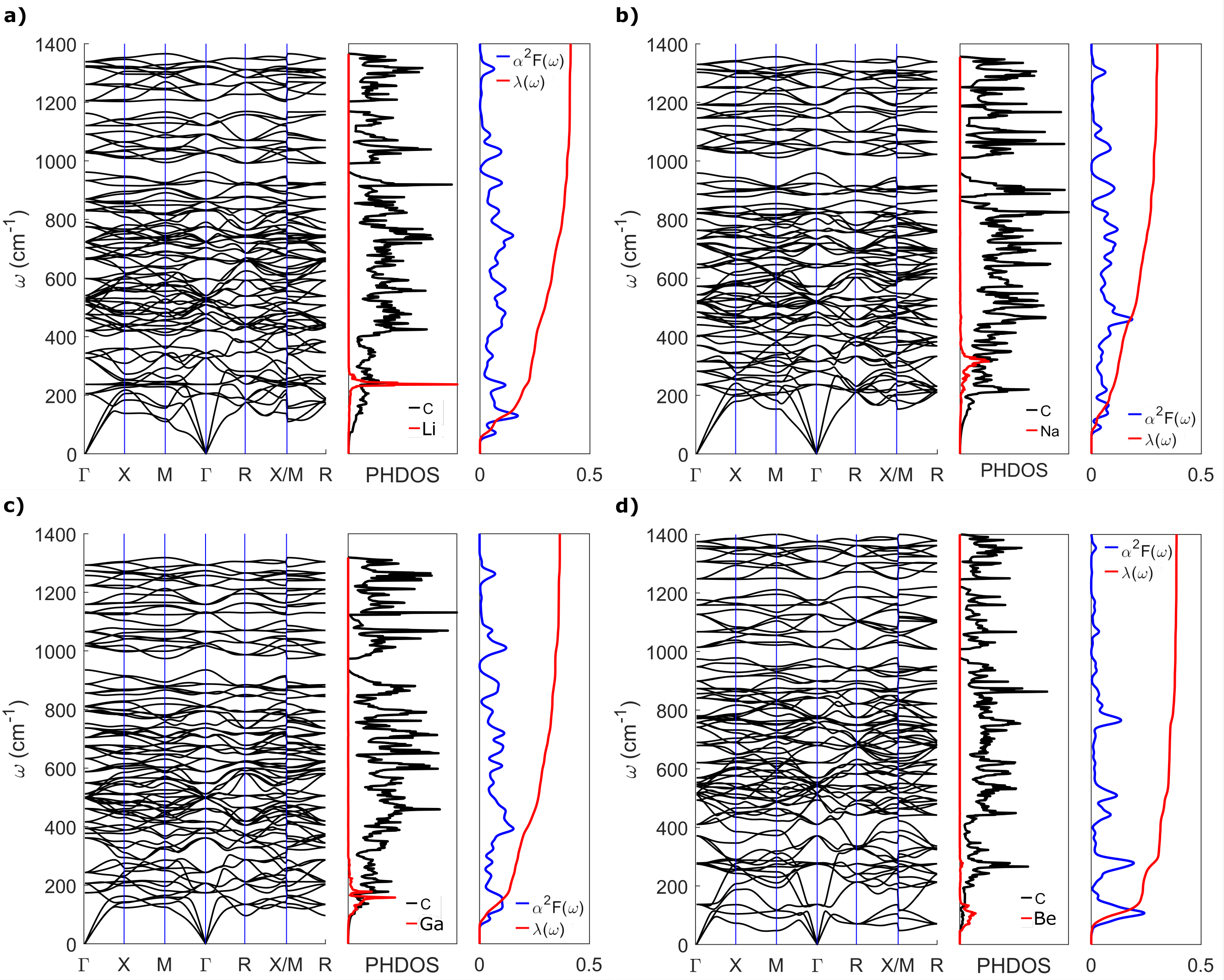}
	\caption{Phonon dispersion curves (left panels), projected phonon density of states (middle panels), Eliashberg spectral function $\alpha^2 F(\omega)$, blue line, and integrated electron-phonon coupling $\lambda(\omega)$, red line, (right panels) for the doped clathrate structures: a) Li@$\mathrm{C}_{24}$; b) Na@$\mathrm{C}_{24}$; c) Ga@$\mathrm{C}_{24}$; d) Be@$\mathrm{C}_{60}$.}
	\label{fig_lamb2}
\end{figure*} 

Given the unfavorable results just described, the search for superconductivity was pursued through a systematic search of doped clathrate structures employing dopants from groups I, II and III  (Li, Na, K, Be, Mg, Ca, Al, Ga) and also from group IV (Ge). As already mentioned, in the doped clathrates considered, the guest atoms are inserted in only one of the two cages available in the structure. The electronic bands, DOS and phonon dispersion curves for all the doped clathrates being studied are given in the SI, sections D and E, respectively. The electronic structures across the M@$\mathrm{C}_{60}$ series essentially follow a rigid-band model, in particular for the monovalent doped structures, the guest atom donates most of its charge to the carbon framework (see section D of the SI, Figures S3 and S4). The most noticeable exceptions are Be@$\mathrm{C}_{60}$ and Mg@$\mathrm{C}_{60}$, shown in Figure S3, where the metal states hybridize with the carbon states, thus changing deeply the electronic structure around the Fermi level. An interesting case is Ca@$\mathrm{C}_{60}$ where the charge transfer shifts the Fermi level to the DOS peak, inducing a high density of states (see the last panel of Figure S3). Concerning the electronic structure of the M@$\mathrm{C}_{24}$ series (also shown in section D of the SI, Figures S5 and S6) the rigid-band model is not followed, with the exceptions of the monovalent doped compounds (M=Li,Na,K) and Ca@$\mathrm{C}_{24}$. The insertion of the dopant in the smaller $\mathrm{C}_{24}$ cage leads to the hybridization of the dopant and carbon framework states changing deeply the band structure.

The phonon dispersion curves for the studied doped fullerite clathrates are shown in the SI, section E. Only four doped clathrate structures do not display imaginary phonon modes and are dynamically stable, these structures being Li@$\mathrm{C}_{24}$, Na@$\mathrm{C}_{24}$, Ga@$\mathrm{C}_{24}$ and Be@$\mathrm{C}_{60}$. The imaginary phonon modes of unstable doped clathrates are either originated from the carbon framework or from mixed carbon-dopant vibrations. In the first case, the charge transfer from the guest atom to the carbon framework structure is observed to be nearly completed, while it is clearly uncompleted in the latter case.

Superconducting behavior was explored for the four stable clathrates. As referred above, the Be@$\mathrm{C}_{60}$ clathrate shows a very different electronic band structure from that of the pristine fullerite clathrate (see Figure \ref{fig_lamb} d). In particular, the band that crosses the Fermi level is the result from the hybridization of beryllium and carbon framework states. Indeed, the incomplete charge transfer from the dopant to the carbon structure is confirmed from the integrated local DOS contours and the projected DOS (PDOS) around the Fermi level (see Figures S11 and S12). The PDOS curve for the Be@$\mathrm{C}_{60}$ shows electronic density from the dopant at the Fermi level, while no density is observed for the Li@$\mathrm{C}_{24}$, Na@$\mathrm{C}_{24}$, indicating in the latter case a complete charge transfer to the framework carbon structure. Ga@$\mathrm{C}_{24}$ presents an intermediate case where $\sim 1$ electron is donated to the carbon framework but there is still a very low amount of states due to the dopant at the Fermi level.
This picture is corroborated by the computed Bader charges given in Table S3. 

\begin{figure}[ht]
	\centering
	\includegraphics[scale=0.62]{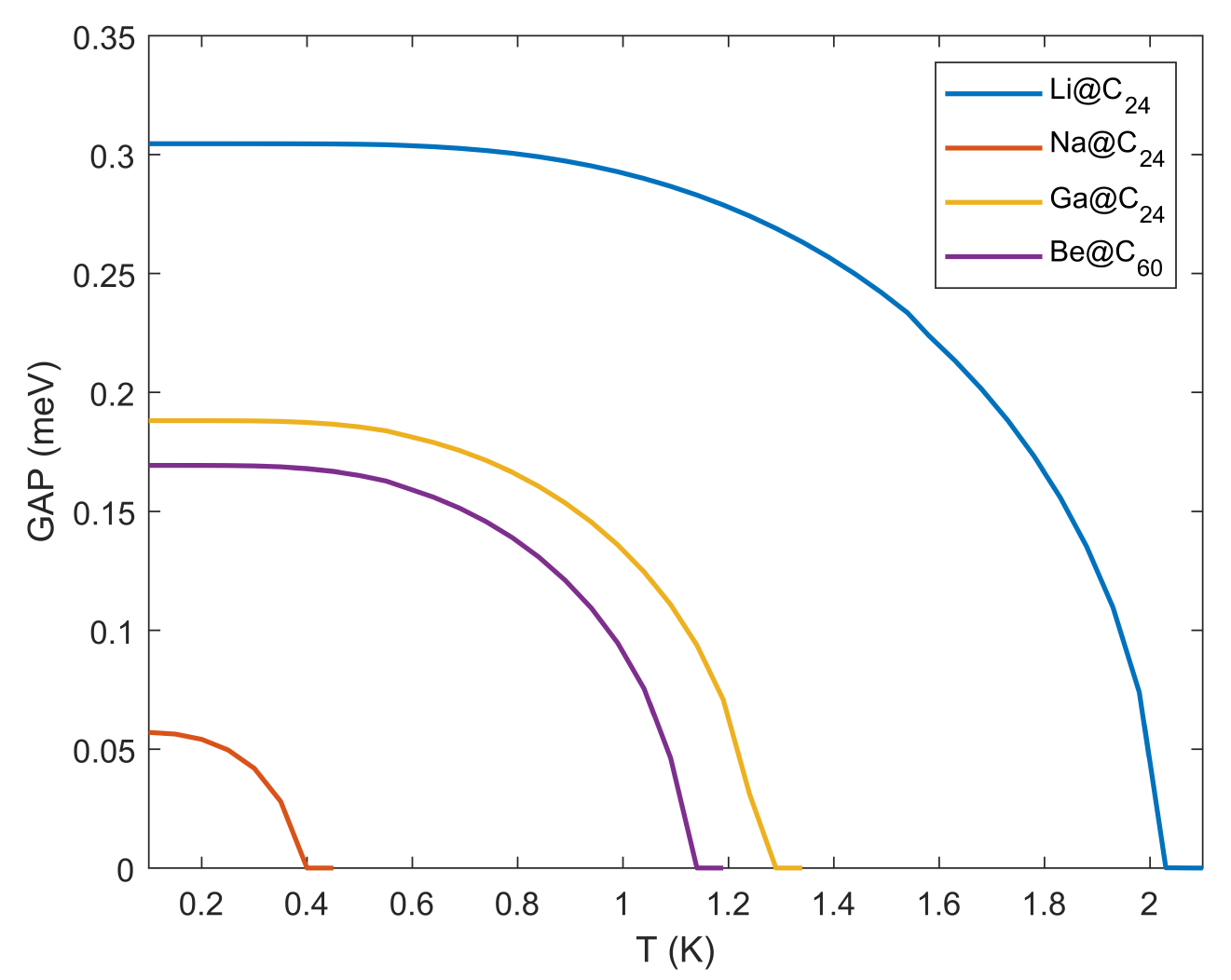}
	\caption{Superconducting gap for the doped $\mathrm{C}_{60}$ clathrate structure: Li@$\mathrm{C}_{24}$ (blue curve); Na@$\mathrm{C}_{24}$ (orange curve); Ga@$\mathrm{C}_{24}$ (yellow curve); Be@$\mathrm{C}_{60}$ (purple curve).}
	\label{fig_tc}
\end{figure}

The phonon dispersion curves shown in the left panels of Figure \ref{fig_lamb2} present two distinct zones separated around 1000 cm$^{-1}$. A second gap around 1200 cm$^{-1}$ is also observed for Li@$\mathrm{C}_{24}$ and Be@$\mathrm{C}_{60}$ compounds. The middle panels in Figure \ref{fig_lamb2} present the projected phonon density of states (PHDOS) showing that the dopant atoms contribute only to lower frequencies, below 400 cm$^{-1}$. The calculated Eliashberg spectral function, $\alpha^2$F$(\omega)$, and the electron-phonon coupling constant, $\lambda (\omega)$, are also shown in Figure \ref{fig_lamb2} (right panels). The strongest peaks in the Eliashberg spectral function arise at low energy, between 100 and 300 cm$^{-1}$, for Be@$\mathrm{C}_{60}$ and thus these are the phonon frequencies that will mostly contribute to the electron-phonon coupling ($\sim80\%$) and concomitantly to the superconducting T$_c$, see Figure \ref{fig_lamb2} d). For the other systems, the contribution to the Eliashberg spectral function is more evenly distributed over all frequencies leading to a smother growth of the integrated electron-phonon coupling curve, Figures \ref{fig_lamb2} a), b) and c). Hence, beryllium is the only dopant that contributes significantly to the Eliashberg spectral function, wich is in agreement with the fact that it is the sole dopant whose electronic states strongly hybridize with the carbon states. The integrated electron-phonon coupling constant is always lower than 0.5, as indicated in Table \ref{at3}, considerably smaller than the 2.92 reported for the sodium doped sodalite-like carbon clathrate \cite{sodalite}, and thus the T$_c$'s should not be high. Indeed, using the Allen-Dynes modified McMillan equation with a Coulomb screening potential $\mu^*$=0.1 the T$_c$'s obtained are quite low: 1.81 K, 0.23 K, 0.99 K and 0.88 K, respectively for Li@$\mathrm{C}_{24}$, Na@$\mathrm{C}_{24}$, Ga@$\mathrm{C}_{24}$ and Be@$\mathrm{C}_{60}$ clathrates. We also solved the isotropic Eliashberg equations obtaining the superconducting gaps for all the stable structures, shown in Figure \ref{fig_tc}. The calculated T$_c$'s are 2.03 K, 0.40 K, 1.29 K and 1.14 K for Li@$\mathrm{C}_{24}$, Na@$\mathrm{C}_{24}$, Ga@$\mathrm{C}_{24}$ and Be@$\mathrm{C}_{60}$, respectively, in agreement with those referred above.   


The electron-phonon coupling constants, for each high-symmetry q-point, are given in Table \ref{at3}. The Na@$\mathrm{C}_{60}$ clathrate has been included in this list because, despite being unstable, it has a higher electron-phonon coupling constant than all the stable doped clathrates at X and M q-points ( excluding the X point for Be@$\mathrm{C}_{60}$ clathrate). In addition, it has, by far, the highest density of states at the Fermi level, N(E$_F$). If this unstable structure could be turned dynamically stable, by accounting anharmonic \cite{LaH10_Ion,ion_nature,ion_prl} and/or pressure effects a serious increase in the T$_c$ would then be expected. Considering that its softest phonon is barely imaginary, $56i$ cm$^{-1}$, it is possible that such effects will be sufficient to turn the structure stable, as was observed for the superconductor LaH$_{10}$ \cite{LaH10_Ion} and for the ferroelectric SnTe \cite{SnTe_Ion}. 


\begin{table}[H]
	\centering
	\caption{Superconducting properties of a few selected systems. The electron-phonon coupling constant at R q-point is not present for Na@C$_{60}$ due to the existence of imaginary modes. The number of density of states at the Fermi level N(E$_F$) is given in state per spin per meV and per atom in the last column.} 
	\label{at3}
	\resizebox{\columnwidth}{!}{\begin{tabular}{c | c | c | c | c | c | c | c | c } 
			            & $\lambda$    & $\lambda$    & $\lambda$  & $\lambda$ & $\omega_{log}$ & T$_C$ & N(E$_F$)\\
               & X q-point & M q-point  & R q-point  & integrated & cm$^{-1}$  & K & state/spin/eV/atom\\
			
			\hline 
			&                         &                       &   &   & & & \\
			Li@$\mathrm{C}_{24}$   & 0.3740                  & 0.3368                 & 0.2801                 & 0.41 & 254.69 & 1.81 & 63.0 \\
			Na@$\mathrm{C}_{24}$   & 0.3170                  & 0.2326                 & 0.1597                 & 0.30 & 342.87 & 0.23 & 63.3 \\
			Ga@$\mathrm{C}_{24}$   & 0.3716                  & 0.2263                 & 0.3158                 & 0.36 & 289.45 & 0.99 & 56.4 \\ 
			Be@$\mathrm{C}_{60}$   & 0.9176                  & 0.1835                 &   0                    & 0.39 & 178.34 & 0.88 & 21.4 \\
			Na@$\mathrm{C}_{60}$   & 0.6142                  & 0.7207                 & - & -  & - & - & 105.7 \\
	
	\end{tabular}}
\end{table} 
 
\section*{Conclusion}

The stability and superconducting properties of the fullerite clathrate structures doped with simple metals has been explored. Only four structures were found to be stable: Li@$\mathrm{C}_{24}$, Na@$\mathrm{C}_{24}$, Ga@$\mathrm{C}_{24}$ and Be@$\mathrm{C}_{60}$. They all show superconducting critical temperatures bellow 4 K, due to the weak electron-phonon coupling and the small number of states at the Fermi level. These two parameters, particularly the latter, are higher in the barely unstable Na@$\mathrm{C}_{60}$, indicating that the stabilization of this compound would result in a superconductor with T$_c$ much higher than those reported here. We are currently exploring this possibility by explicitly considering anharmonic and pressure effects.   

\begin{acknowledgments}
%
%
This work was developed within the scope of the project CICECO-Aveiro Institute of Materials, UIDB/50011/2020, UIDP/50011/2020 \& LA/P/0006/2020, financed by national funds through the FCT/MCTES (PIDDAC) and IF/00894/2015 finances by FCT. J. Laranjeira acknowledges a PhD grant from FCT (SFRH/BD/139327/2018). 

I.E. and \DJ.D. acknowledge funding from the European Research Council (ERC) under the European Union’s Horizon 2020 research and innovation program (Grant Agreement No. 802533) and the Department of Education, Universities and Research of the Eusko Jaurlaritza and the University of the Basque Country UPV/EHU (Grant No. IT1527-22).

The work has been performed under the project HPC-EUROPA3 (INFRAIA-2016-1-730897), with the support of the EC Research Innovation Action under the H2020 Programme; in particular, the author gratefully acknowledges the support of the computer resources and technical support provided by BSC.
\end{acknowledgments}

\bibliography{referencias2.bib}

\end{document}


\preprint{APS/123-QED}
\onecolumngrid
\section*{Supporting Information}
\section*{A. Electronic Structure of the undoped fullerite clathrate   }
 \onecolumngrid
  \begin{figure}[H]
		\hspace*{-1.5cm} 
	\centering
	\includegraphics[scale=0.65]{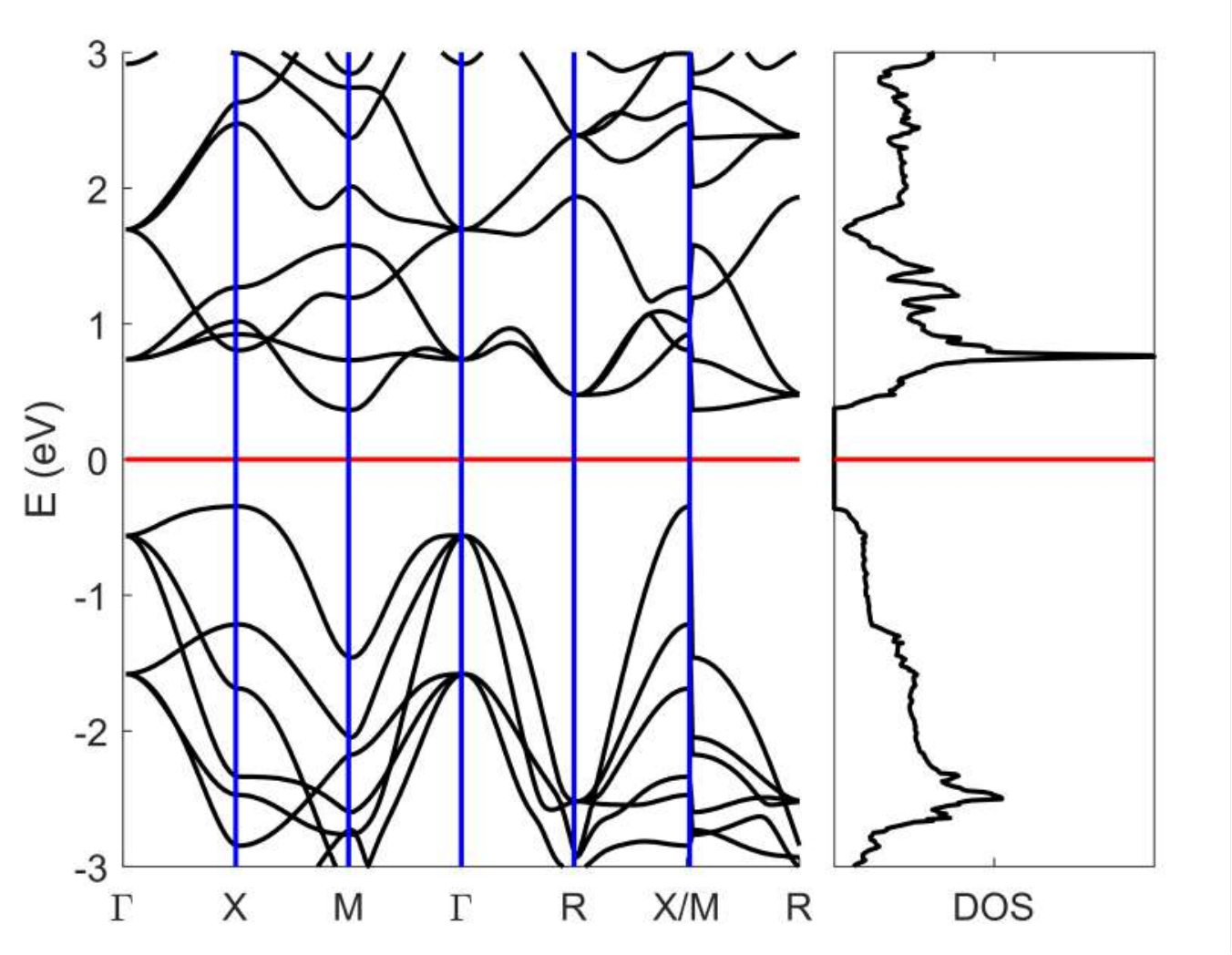}
	\caption{Electronic density of states (right panel) and electronic band structure (left panel) of the fullerite C$_{60}$ clathrate. Red lines indicate the Fermi level.}
	\label{b3lyp}
\end{figure}

\section*{B. Energy and Lattice parameter evolution with dopant insertion }
 
\begin{longtable}{p{2cm} | p{2cm} | p{2cm}  }
 M(dopant)  & @C$_{60}$  &  @C$_{24}$ \\
 	\hline
 Li         &  6.2211	  &  6.2358   \\
 Na         &  6.2258	  &  6.2565   \\
 K          &  6.2366	  &  6.3039   \\
 Be         &  6.2067	  &  6.2544   \\
 Mg         &  6.2187	  &  6.2824   \\
 Ca         &  6.2359	  &  6.2990   \\
 Al         &  6.2203	  &  6.2872   \\
 Ga         &  6.2200	  &  6.2926   \\  
 Ge         &  6.2111	  &  6.3087   \\  
 	\caption*{Table S1. Simple cubic lattice parameters for all the doped systems after optimization, given in \AA. The undoped system presents a lattice parameter of 6.2031 \AA.}
 	\label{s1}
 \end{longtable} 

\begin{longtable}{p{2cm} | p{2cm} | p{2cm}  }
 M(dopant)  & @C$_{60}$  &  @C$_{24}$ \\
 	\hline
 Li         & -7.5209 & -7.4616 \\
 Na         & -7.5161 & -7.4089 \\
 K          & -7.5200 & -7.3202 \\
 Be         & -7.4772 & -7.3950 \\
 Mg         & -7.4666 & -7.3312 \\
 Ca         & -7.5256 & -7.3700 \\
 Al         & -7.5179 & -7.3455 \\
 Ga         & -7.5189 & -7.3270 \\  
 Ge         & -7.5256 & -7.3076 \\  
 	\caption*{Table S2. DFT formation energy for each structure in eV/atom. The formation energy, E$_f$, is computed by subtracting the energy per atom from the structure energy, E$_f$=E$_{stru}$-(30E$_c$+E$_M$). The undoped system presents a ground state energy of -7.7327 eV/atom.}
 	\label{s2}
 \end{longtable}

\section*{C. Electronic structure and phonon dispersion curves of the Sc@C$_{60}$ and Y@C$_{60}$ compounds}

 \begin{figure}[ht!]
		\hspace*{-1.5cm} 
	\centering
	\includegraphics[scale=0.75]{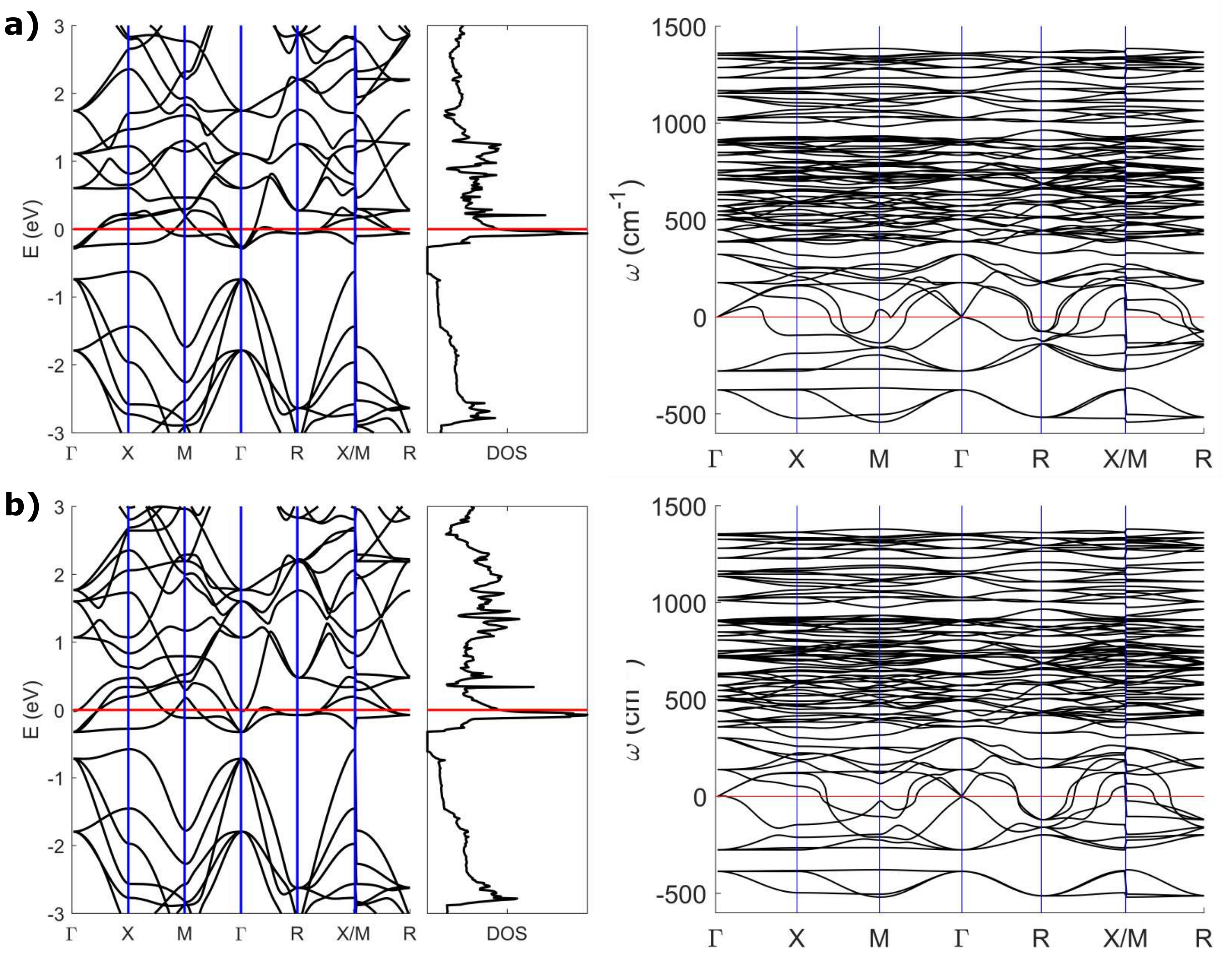}
	\caption{Electronic structure and phonon dispersion curves of fullerite C$_{60}$ clathrate structure doped with two trivalent dopants: a) Sc@$_{60}$; b) Y@$_{60}$. The left panels show the band structure and the electronic DOS, and the right panels show the phonon dispersion curves.}
	\label{s2}
\end{figure}

\section*{D. Electronic Structures  of the doped fullerite clathrates}

\begin{figure}[H]
		\hspace*{-1.5cm} 
	\centering
	\includegraphics[scale=0.65]{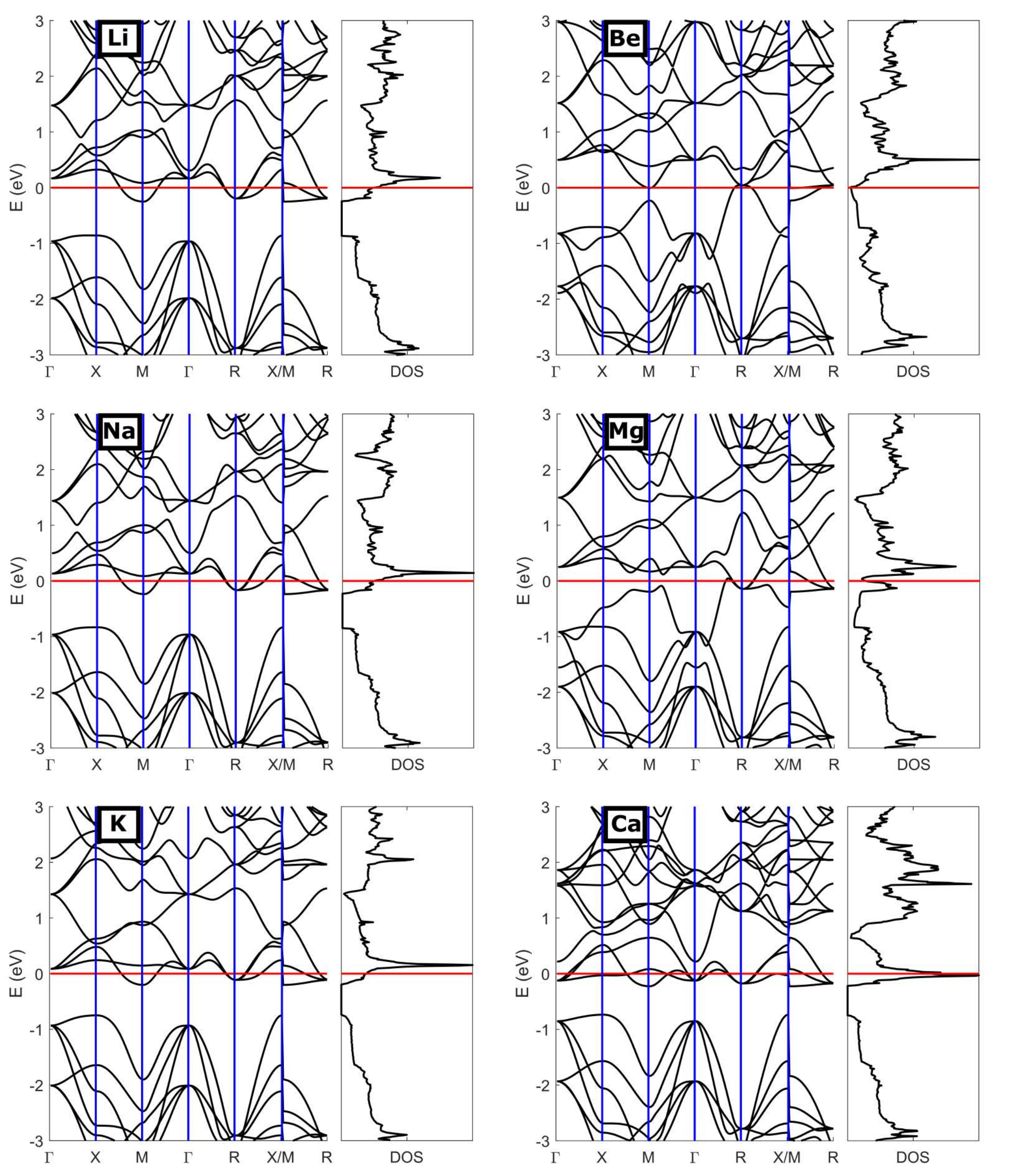}
	\caption{Electronic band structure and density of states of the doped fullerite clathrates containing the guest-dopant in the C$_{60}$ cages. The left columns correspond to the monovalent dopants and the right columns correspond to the divalent ones.}
	\label{S3}
\end{figure}

\begin{figure}[H]
		\hspace*{-1.5cm} 
	\centering
	\includegraphics[scale=0.65]{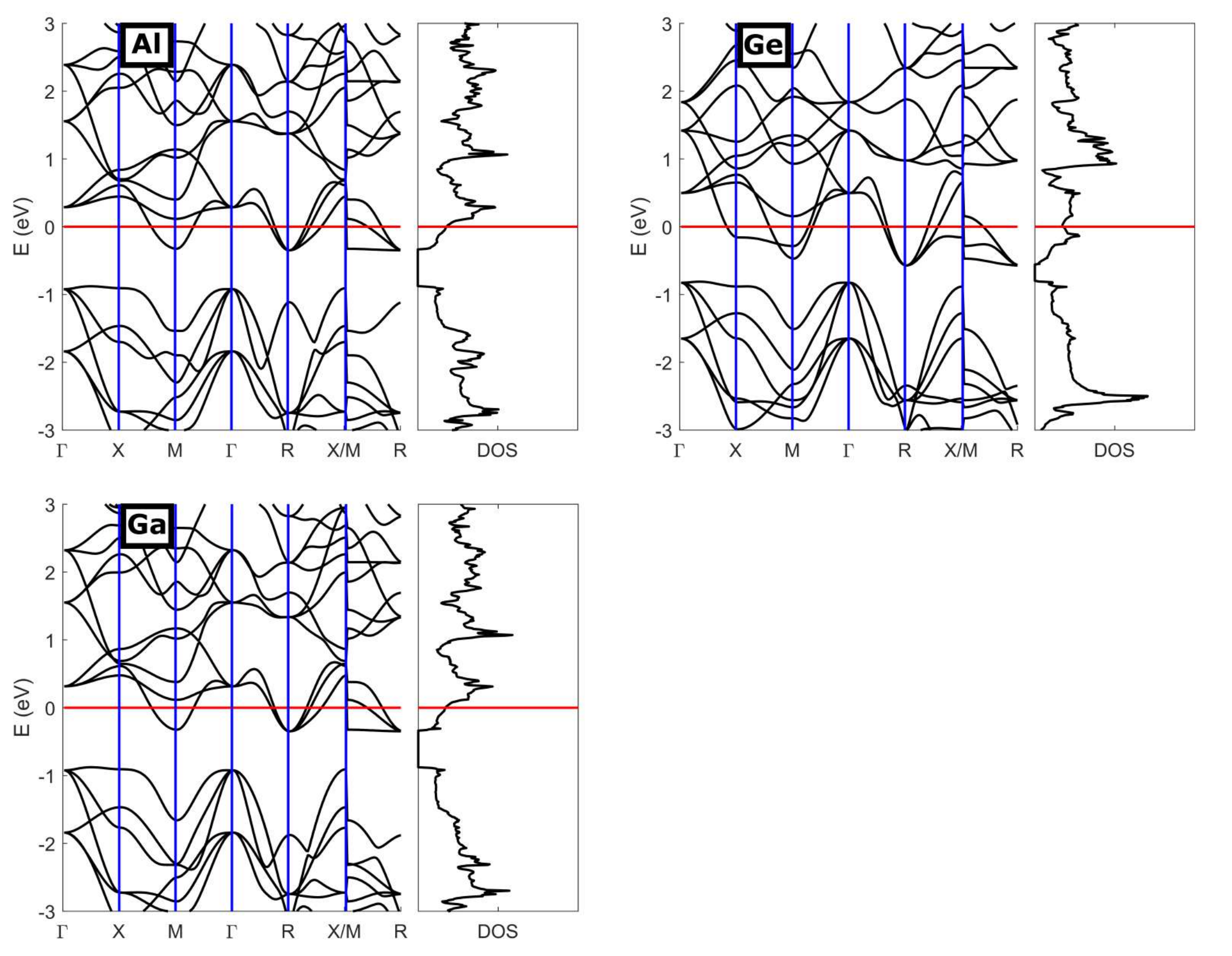}
	\caption{Electronic band structure and density of states of the doped fullerite clathrate containing the trivalent (left columns) or the tetravalent guest-dopants (right columns) in the C$_{60}$ cages. }
	\label{S4}
\end{figure}

\begin{figure}[H]
		\hspace*{-1.5cm} 
	\centering
	\includegraphics[scale=0.65]{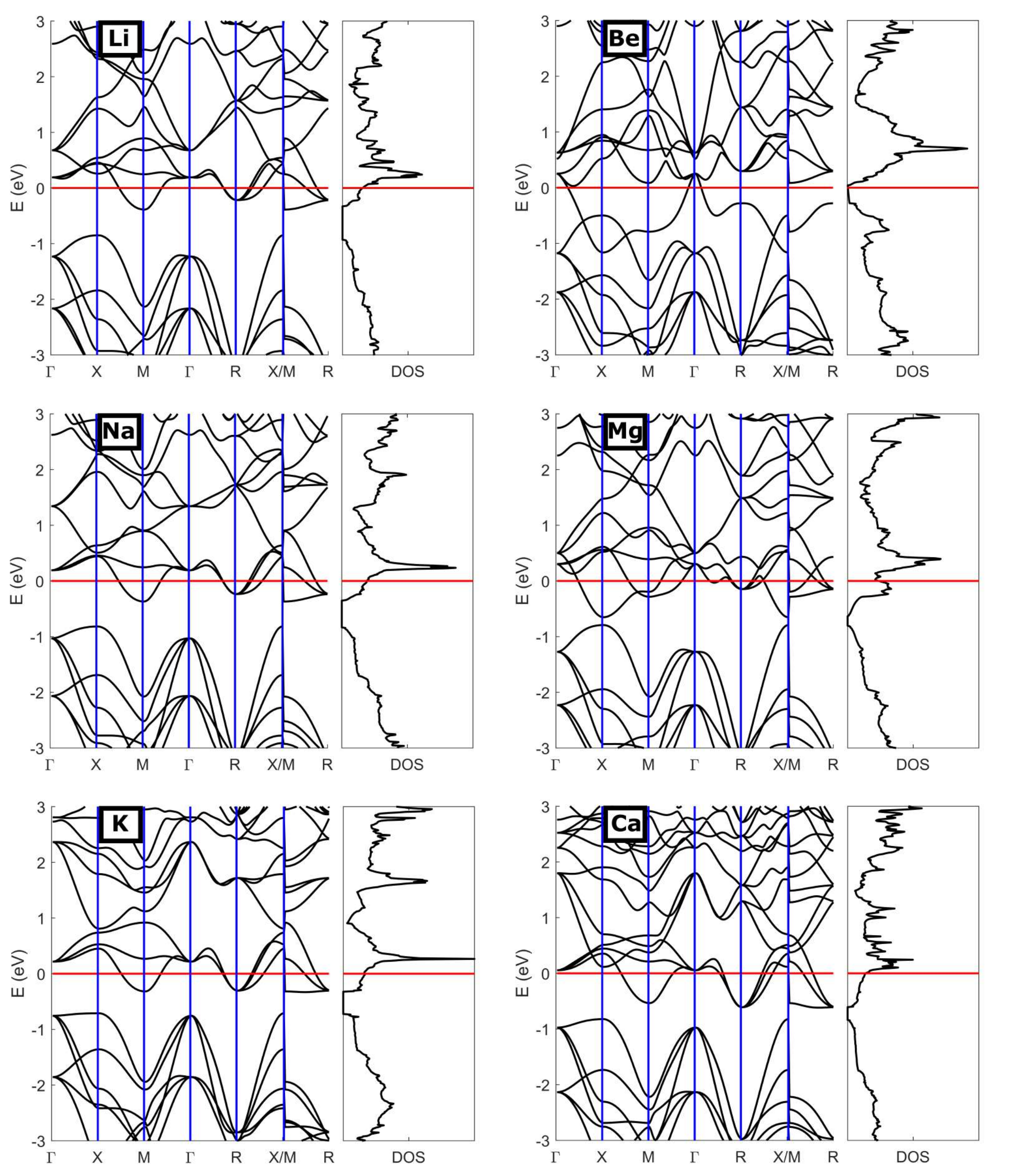}
        \caption{Electronic band structure and density of states of the doped fullerite clathrates containing the guest-dopants in the C$_{24}$ cages. The left columns correspond to the monovalent dopants and the right columns correspond to the divalent dopants.}
	\label{S5}
\end{figure}

\begin{figure}[H]
		\hspace*{-1.5cm} 
	\centering
	\includegraphics[scale=0.65]{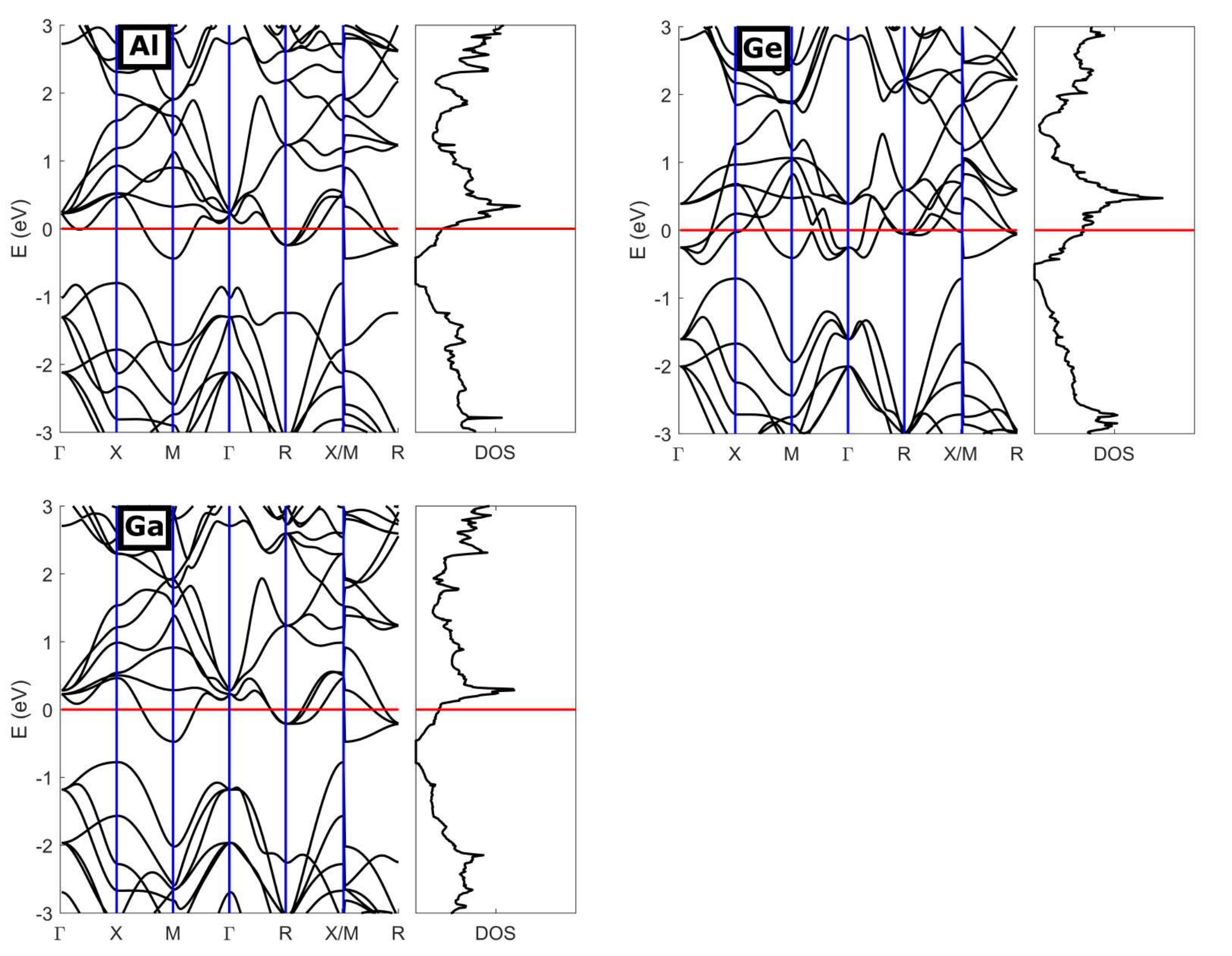}
        \caption{Electronic band structure and density of states of the doped fullerite clathrates containing the trivalent  (left columns) or tetravalent (right columns) guest-dopants in the C$_{24}$ cages. }
	\label{S6}
\end{figure}

\section*{E. Phonon Dispersion Curves for the doped fullerite clathrates}

 \begin{figure}[H]
		\hspace*{-1.5cm} 
	\centering
	\includegraphics[scale=0.65]{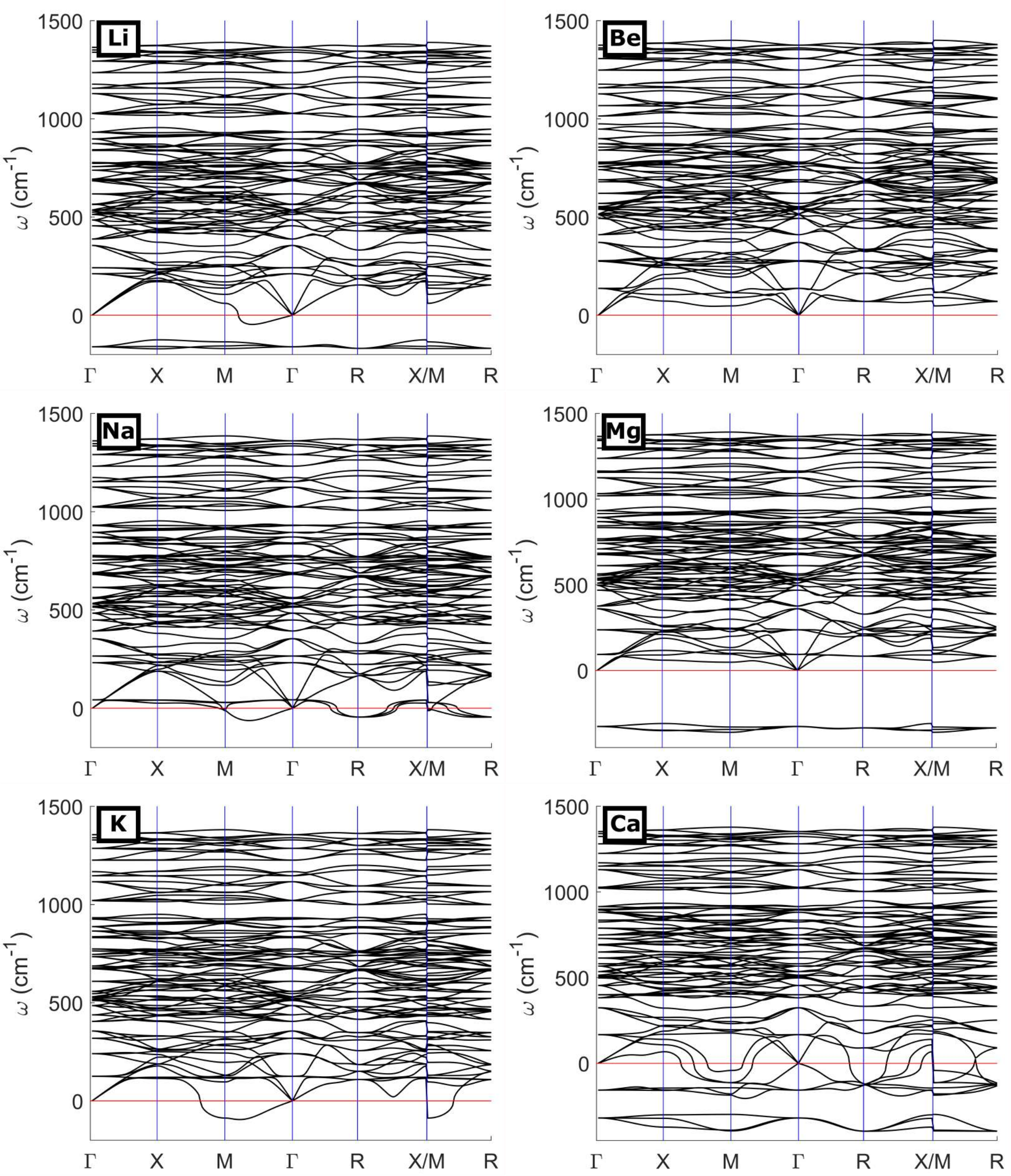}
	\caption{Phonon spectra of the doped fullerite clathrates containing the monovalent (left column) or divalent (right column) guest-dopants in the C$_{60}$ cages. }
	\label{S7}
\end{figure}

 \begin{figure}[H]
		\hspace*{-1.5cm} 
	\centering
	\includegraphics[scale=0.65]{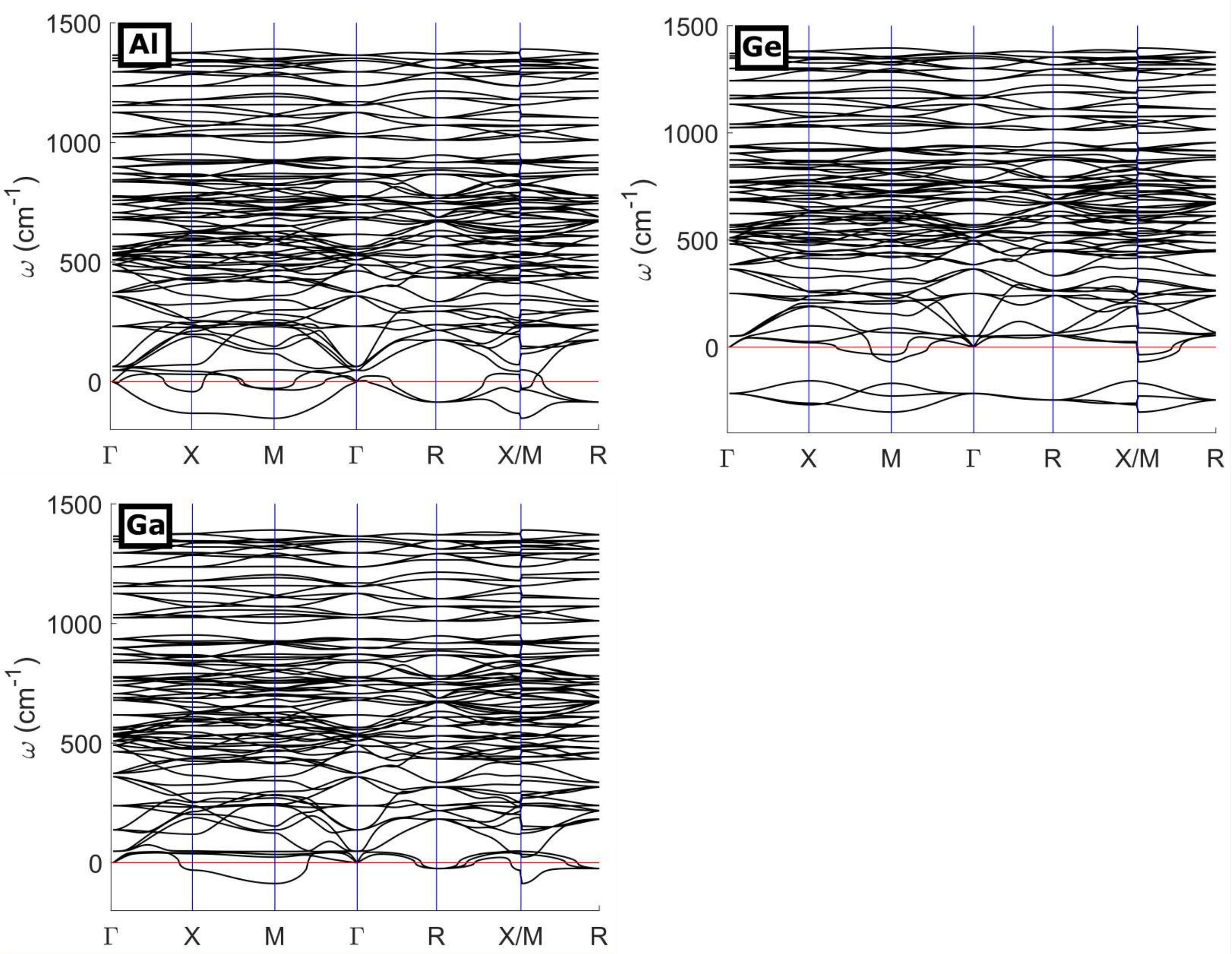}
	\caption{: Phonon spectra of the doped fullerite clathrates containing the trivalent (left column) or tetravalent (right column) guest-dopants in the C$_{60}$ cages.}
	\label{S8}
\end{figure}

 \begin{figure}[H]
		\hspace*{-1.5cm} 
	\centering
	\includegraphics[scale=0.65]{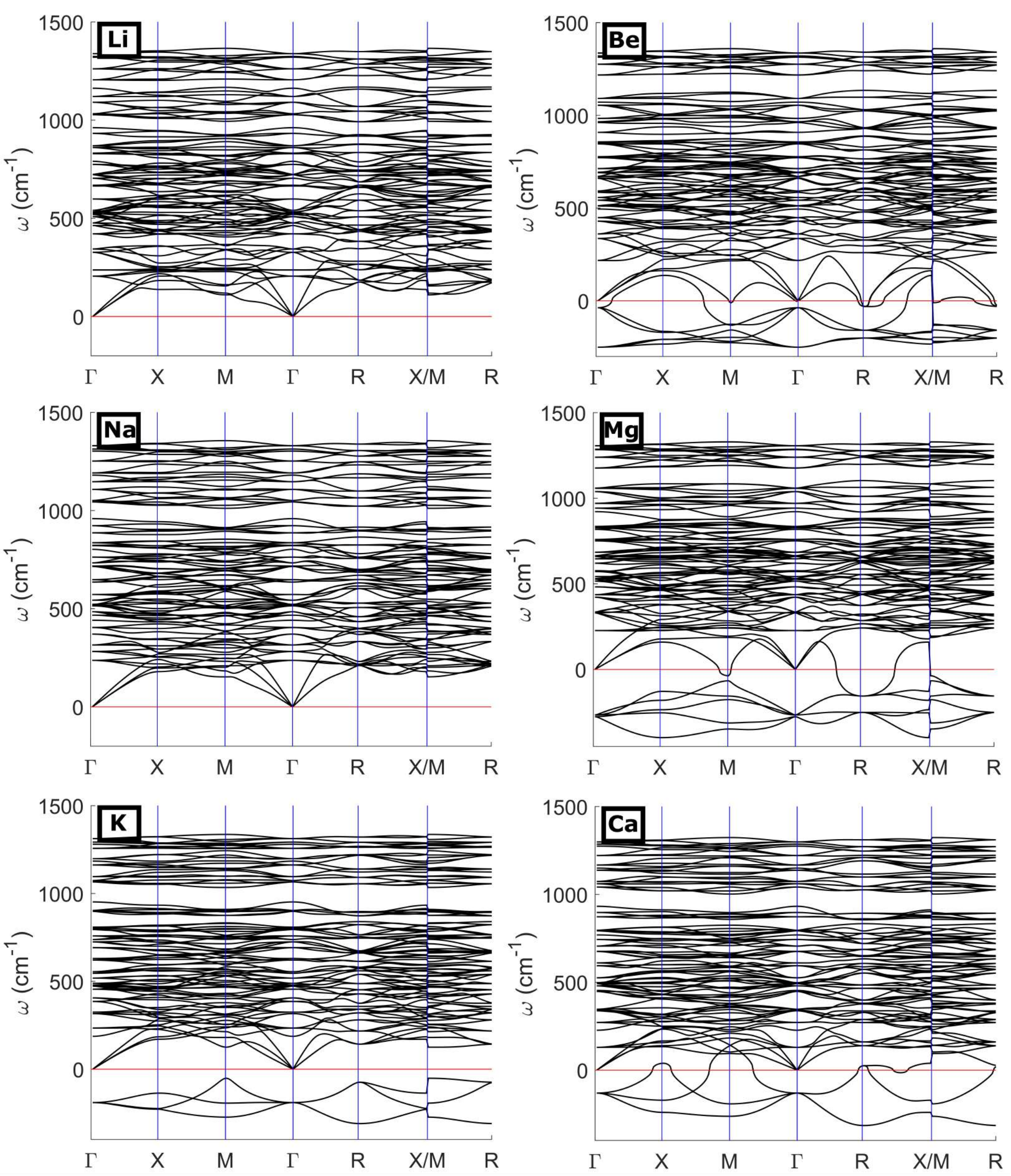}
	\caption{Phonon spectra of the doped fullerite clathrates containing the monovalent (left column) or divalent (right column) guest-dopants in the C$_{24}$ cages.}
	\label{S9}
\end{figure}

 \begin{figure}[H]
		\hspace*{-1.5cm} 
	\centering
	\includegraphics[scale=0.65]{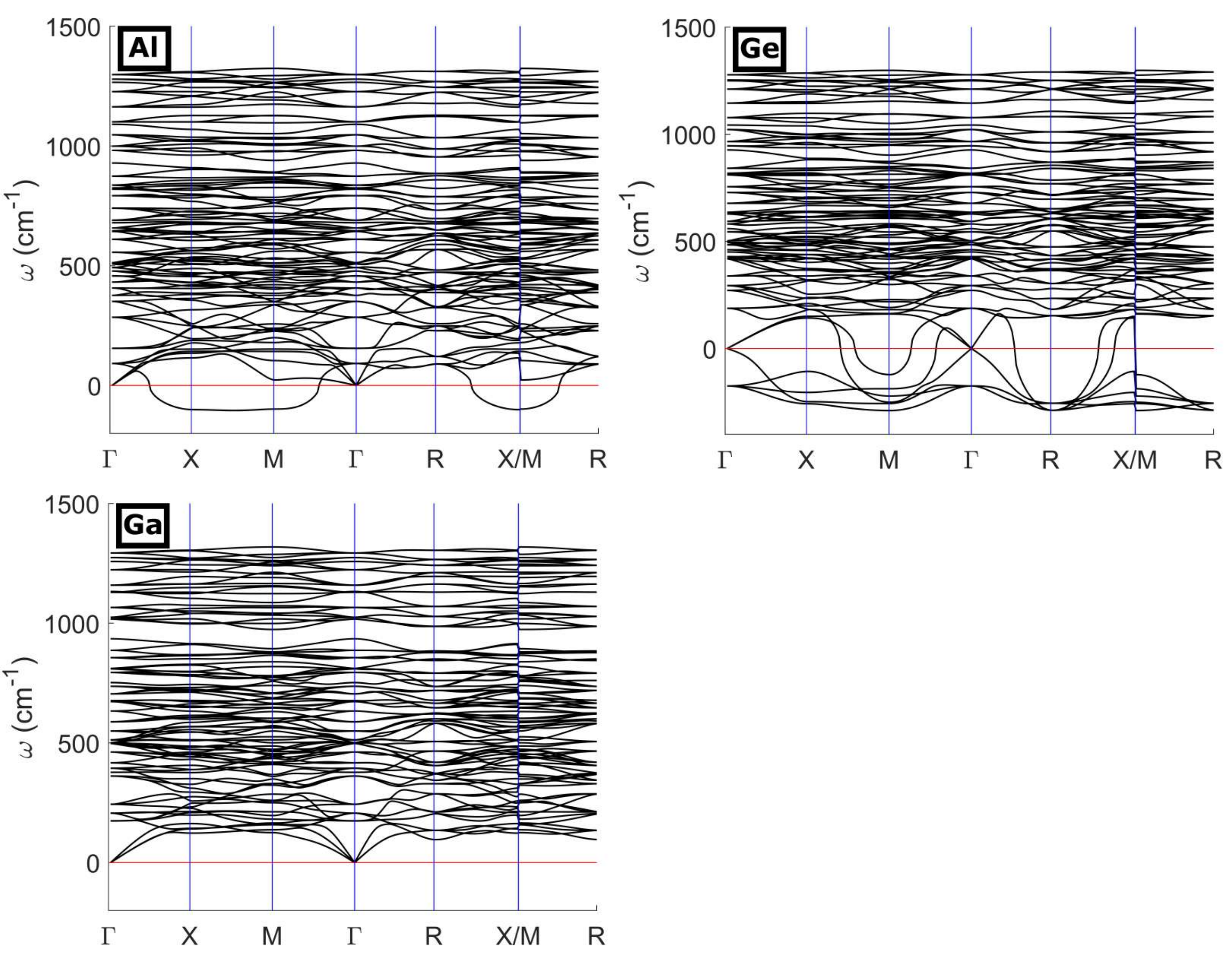}
	\caption{Phonon spectra of the doped fullerite clathrates containing the trivalent (left column) or tetravalent (right column) guest-dopants in the C$_{24}$ cages.}
	\label{S10}
\end{figure}

\clearpage

\section*{F. Integrated local electronic density around the Fermi level for the stable doped fullerite clathrates}
  
 \begin{figure}[ht!]
		\hspace*{-1.5cm} 
	\centering
	\includegraphics[scale=0.25]{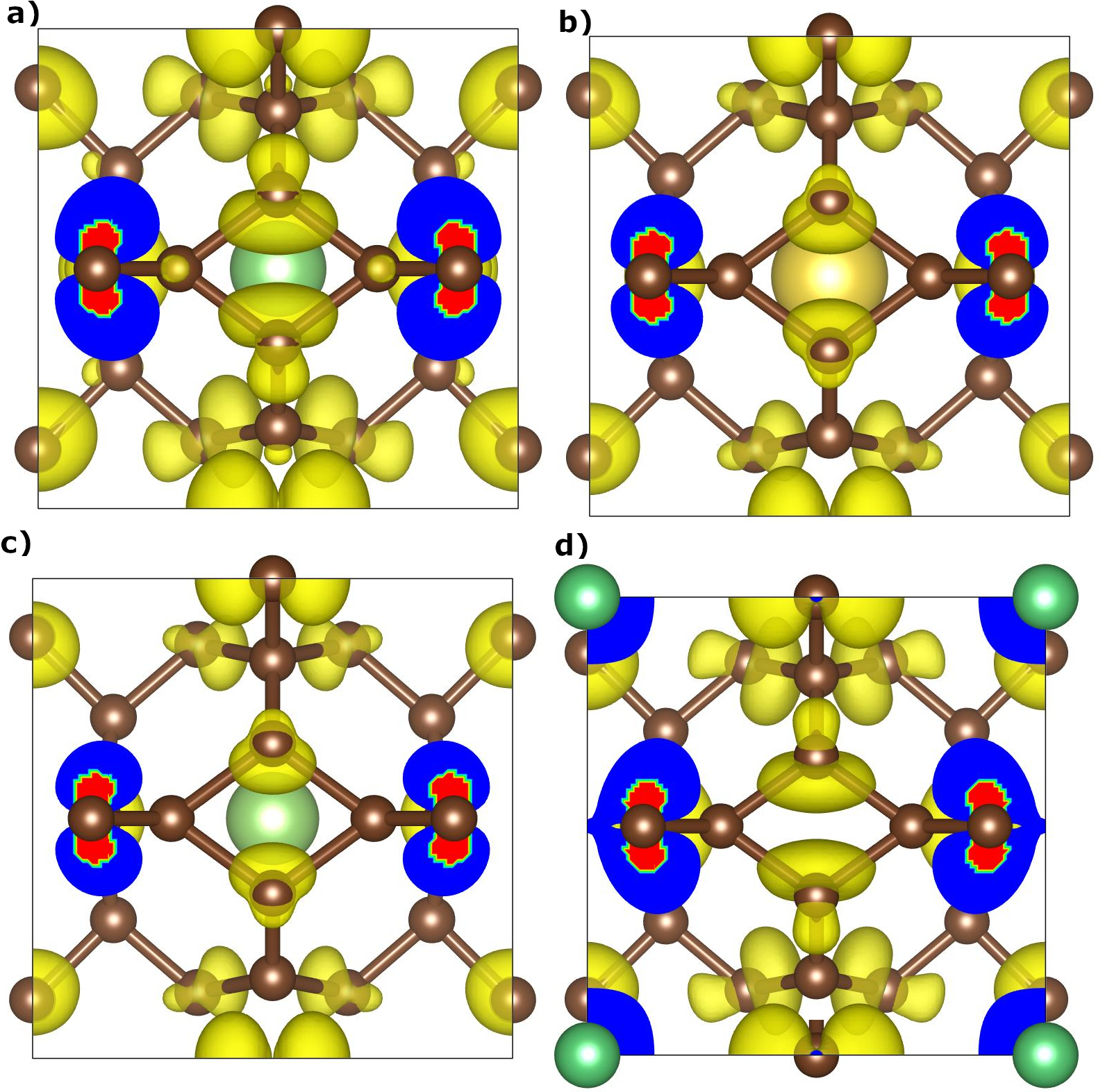}
	\caption{Integrated local electronic density around the Fermi level for:a) Li@C$_{24}$; b) Na@C$_{24}$; c) Ga@C$_{24}$ and d) Be@C$_{60}$. The partial electron density was integrated 0.01 eV around the Fermi level, with the partial electronic density isosurface with $2.0\times10^{-4}$ e/\AA$^3$ ($5.0\times10^{-5}$ e/\AA$^3$  for Be@C$_{60}$) superposed on a ball-and-stick model for the (100) crystallographic plane.}
	\label{S11}
\end{figure}
\clearpage

\section*{G. Projected electronic density of states for the stable doped clathrates}
 \begin{figure}[H]
		\hspace*{-1.5cm} 
	\centering
	\includegraphics[scale=0.57]{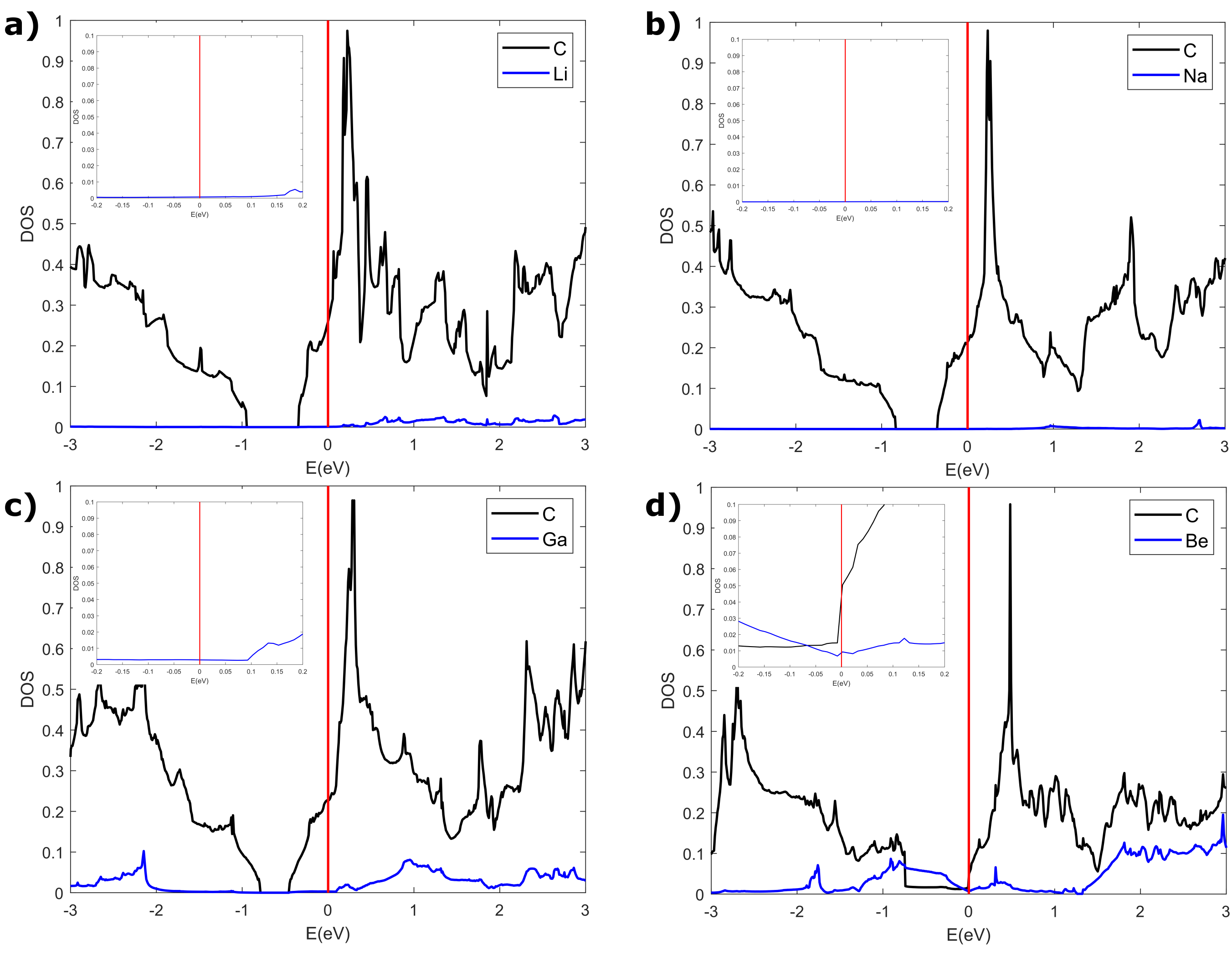}
	\caption{Projected electronic density of states for: a) Li@C$_{24}$; b) Na@C$_{24}$; c) Ga@C$_{24}$ and d) Be@C$_{60}$. The black always denotes carbon projected DOS while blue denotes the different dopant. Fermi level is denoted by a red line. All the plots are normalized to the highest peak found in carbon projected DOS.}
	\label{S12}
\end{figure}

\section*{H. Bader charges for the stable structures}
\begin{longtable}{p{2cm} | p{2cm} | p{2cm}| p{2cm}  }
 Structure & Atom & Wyckoff Position &	Bader charge \\
 	\hline
            & C1 & 12k	&  4.05 \\
Li@$C_{24}$ & C2 & 12k	&  3.99 \\
            & C3 & 6f	&  4.07 \\
            & Li & 1b	&  0.11 \\
      \hline
            & C1 & 12k	&  4.05 \\
Na@$C_{24}$ & C2 & 12k	&  3.99 \\
            & C3 & 6f	&  4.06 \\
            & Na & 1b	&  0.16 \\
  \hline
            & C1 & 12k	&  4.06 \\
Ga@$C_{24}$ & C2 & 12k	&  3.99 \\
            & C3 & 6f	&  4.06 \\
            & Ga & 1b	&  2.08 \\
              \hline
            & C1 & 12k	&  4.02 \\
Be@$C_{60}$ & C2 & 12k	&  3.95 \\
            & C3 & 6f	&  4.08 \\
            & Be & 1a	&  1.92 \\
 	\caption*{Table S3. Bader charges of the inequivalent atoms. }
 	\label{ts3}
 \end{longtable}